\documentclass[10pt,letterpaper,twocolumn]{article} 

\usepackage{ol2}
\usepackage[draft]{hyperref}
\usepackage{amsmath}

\begin{document}

\twocolumn[ 

\title{A Monolithic Filter Cavity for Experiments in Quantum Optics}

\author{Pantita Palittapongarnpim, Andrew MacRae, A. I. Lvovsky$^*$}

\address{
Department of Physics and Astronomy, University of Calgary\\
Calgary, Alberta, Canada T2N 1N4\\
$^*$Corresponding author: lvov@ucalgary.ca
}


\begin{abstract}By applying a high-reflectivity dielectric coating on both sides of a commercial plano-convex lens, we produce a stable monolithic Fabry-Perot cavity suitable for use as a narrow band filter in quantum optics experiments. The resonant frequency is selected by means of thermal expansion. Owing to the long term mechanical stability, no optical locking techniques are required. We characterize the cavity performance as an optical filter, obtaining a $45$dB suppression of unwanted modes while maintaining a transmission of 60\%.\end{abstract}

 ] 


\noindent \textbf{Introduction}:

Modern day experiments in quantum optics require the isolation of single-photon level signals from an intense optical background of similar wavelength. Among the many examples there are experiments in quantum optical engineering \cite{Polzik,Furusawa,Andrew}, quantum memory \cite{Gem_11}, and repeaters \cite{vanderwal03}, which require the selection of a specific spectral and spatial mode while simultaneously obtaining high rejection of all other modes. This can be achieved by means of a solid etalon filter or a cavity formed between independent mirrors. However, both these approaches are problematic. In the latter case, active locking of the mirror separation is required, resulting in significant experimental complications. On the other hand, an etalon is intrinsically stable and requires no optical locking, but the planar cavity geometry limits the achievable cavity finesse and provides no spatial transverse-mode filtering \cite{Roy-Hercher}.

Our approach is to combine the intrinsic stability of a monolithic etalon with the high finesse and spatial mode filtering allowed by Fabry-Perot cavities with spherical mirrors by employing a solid plano-convex resonator constructed from a single substrate. This has the advantage of not only long term stability and desirable single-pass suppression of unwanted modes, but also experimental simplicity.


\noindent \textbf{Theory:} The quality of a cavity filter is governed by its finesse which in ideal case is \begin{equation}\label{finesseEq}
\mathcal{F}_R=\pi\sqrt R/(1-R),
\end{equation}
where $R$ is the reflectivity of the mirror surfaces (both mirrors are assumed identical throughout the paper). In reality, the finesse of a cavity can be reduced due to intracavity loss and mismatch between the cavity and incident optical mode:

\begin{equation}
\frac{1}{\mathcal{F}^2_{eff}}=\frac{1}{\mathcal{F}_{R}^2}+\frac{1}{\mathcal{F}_{\rm defect}^2}+\frac{1}{\mathcal{F}_{\rm mode}^2},
\label{effective finesse}
\end{equation}
\noindent where $\mathcal{F}_{R}$ is the ideal finesse, $\mathcal{F}_{\rm defect}$ is the finesse associated with surface defects
 \cite{Sloggett_84}, $\mathcal{F}_{\rm mode}$ is the finesse due to the mismatch of the wavefront and the surface. $\mathcal{F}_{\rm mode}$ can be increased with more precise alignment while $\mathcal{F}_{\rm defect}$ is determined by the optical quality of the cavity substrate.

The effective finesse of a flat-surface etalon is limited by the fact that this cavity configuration is at the border of the stability region. Transverse eigenmodes of the flat-surface cavity have infinite spatial extent. Transverse mode-matching demands that the wavefront matches the cavity mirrors identically, but, if the incident wave is finite, the condition of flat wavefronts cannot be met at both mirrors owing to diffraction, limiting $\mathcal{F}_{\rm mode}$ in  Eq.~(\ref{effective finesse}). This may be partially compensated by choosing a wider beam diameter, but this exposes the beam to a larger range of surface defects which in turn bounds $\mathcal{F}_{\rm defect}$. As a result, the finesse of the flat etalon is limited to approximately 100. To obtain sufficiently high extinction, flat surface filters must employ multi-pass designs or multiple resonators, reducing the maximum transmission at resonance and increasing experimental complexity \cite{Benson}.

A  Fabry-Perot cavity with spherical mirrors is more forgiving in these regards. The focusing induced by the curved mirrors eliminates divergence of the internal field. Moreover, the stability of a spherical Fabry-Perot also translates into less beam wandering \cite{Hernandez} even for a nonparaxial incoupling light which reduces the effect of surface defects, so finesse levels up to the order of $10^5$ can be routinely achieved \cite{maunz2004,hood2001}. Another benefit of this design is that separate transverse modes are non-degenerate. As a result, the cavity provides spatial as well as spectral filtering, which is useful for applications where the signal is not in the same spatial mode as the unwanted background light.

The concave filter cavity is typically implemented using two or more spatially separated mirrors. The stability of the resonant mode is achieved by monitoring the transmission of an auxiliary beam and providing feedback to a piezoelectric transducer on which one of the mirrors is mounted. A drawback of this method is the locking beam adds extra light to the system which must in turn be filtered out. Additionally, care must be taken to minimize residual phase jitter stemming from the locking electronics as well as mechanical lability of the piezo.

A monolithic concave cavity, implemented in this work, combines advantages of both etalons and spherical Fabry-Perot filter cavity: it is mechanically stable and permits high finesse and subsequently high extinction. It can be produced by applying a dielectric coating to a commercial convex lens and is therefore relatively inexpensive. An apparent challenge of this design, associated with tuning the cavity to a particular resonance, can be easily overcome using temperature control. Changing the temperature $T$ of the cavity influences both the cavity length $L$ and index of refraction $n$, resulting in a shift of the resonant frequency $\nu$:
\begin{equation}
	\frac{d\nu}{dT}\approx-\left(\alpha+\frac{1}{n}\frac{\partial n}{\partial T}\right)\nu,
	\label{temperature tunability}
\end{equation}
where $\alpha$ is the thermal expansion coefficient and $\partial n/\partial T$ is obtained from Sellmeyer equations. Here we neglect the change in index of refraction due to frequency as it is much smaller than other relevant factors.

When designing the cavity, one needs to decide upon its length (thickness) $L$, surface curvature radii $r_1,\ r_2$ and reflectivity $R$. These parameters can be calculated from the desired final specifications of the filter, such as the free spectral range, linewidth and extinction ratio, based on the following considerations. Neglecting the defects associated with the imperfect mode matching and surface defects, the transmission of a Fabry-Perot cavity as a function of the frequency is given by \cite{Hecht}
\begin{equation}\label{transmission}
{\mathcal T}=\left(\frac T{1-R}\right)^2\frac 1{1+\frac{4R}{(1-R)^2}\sin^2\frac\delta 2},
\end{equation}
where $\delta$ is the frequency-dependent per-roundtrip phase shift and $T$ is the transmission of the mirror surface (which may not equal to $1-R$ due to losses inside the cavity or at reflection). From the above, we find the transmission at the center of a resonant mode ($\delta=0$) and in the middle between modes ($\delta=\pi$) to equal, respectively
\begin{equation}\label{}
{\mathcal T}_{\rm max}=\left(\frac T{1-R}\right)^2; \quad {\mathcal T}_{\rm min}=\left(\frac T{1-R}\right)^2\frac 1{1+\frac{4R}{(1-R)^2}},
\end{equation}
and hence, assuming $R$ close to unity, the extinction ratio equals
\begin{equation}\label{exratEq}
{\mathcal T}_{\rm max}/{\mathcal T}_{\rm min}\approx 4/(1-R)^2\approx(2{\mathcal F}_R/\pi)^2.
\end{equation}
In this way, the desired extinction ratio determines the requirements on the finesse and surface reflectivity. One must, however, keep in mind the degrading effects associated with losses and surface defects. Choosing the reflectivity too high, such that $T\ll 1-R$ or $\pi/(1-R)<{\mathcal F}_{\rm defect}$, will  degrade the cavity transmission on resonance without significantly improving the extinction ratio.

Once the reflectivity is known, the cavity length is obtained from the desired linewidth $\Delta\nu$  via $\Delta\nu=\frac{\text{FSR}}{\mathcal{F}}$, with $\text{FSR}=c/2nL$ being the free-spectral range.

The choice of the mirror curvature is determined by the spatial filtering requirements. The resonance of Hermite-Gaussian mode $\text{TEM}_{mn}$ is located at \cite{Hercher_68}
\begin{eqnarray}
\nu_{qmn}={\rm FSR}&&\\\nonumber &&\hspace{-2cm}\times\left[ q + \frac{1+m+n}{\pi}\arccos{\sqrt{\left(1-\frac{L}{r_{1}}\right)\left(1-\frac{L}{r_{2}}\right)}} \right]
	\label{Hermite-Gauss}
\end{eqnarray}
where the first term represents the plane-wave resonance condition ($q$ being an integer number defining the longitudinal mode) and the second gives the Gaussian mode correction. A few special cases are of interest here. For a plane cavity, the second term vanishes and all transverse modes are frequency degenerate; no spatial filtering occurs. For a confocal cavity, such that e.g. $r_1=r_2=L$, transverse modes are separated by half-integer number of FSRs. In this case, if the cavity is aligned to the TEM$_{00}$ mode, it will also transmit all even TEM modes; only partial spatial filtering occurs. The advantage of the confocal configuration, however, is that if the cavity is configured to reject a particular frequency in both TEM$_{00}$ and TEM$_{01}$ modes, it can be certain to reject that frequency in all other transverse modes.

In the regime where $L\ll r_1,r_2$ (corresponding to our experiment), adjacent transverse modes are generally nondegenerate and are separated by
\begin{equation}\label{transsepEq}
\Delta\nu_\perp=({\rm FSR}/\pi)\sqrt{L/r_1+L/r_2}.
\end{equation}
In this case, full spatial filtering occurs for a specific frequency provided that $\Delta\nu_\perp$ is larger than the cavity linewidth. Although the cavity may transmit undesired spectral components in other transverse modes, these components can be eliminated by spatial filtering of the TEM$_{00}$ mode transmitted through the cavity.

\begin{figure}[htb]
\centerline{\includegraphics[width=8.5cm]{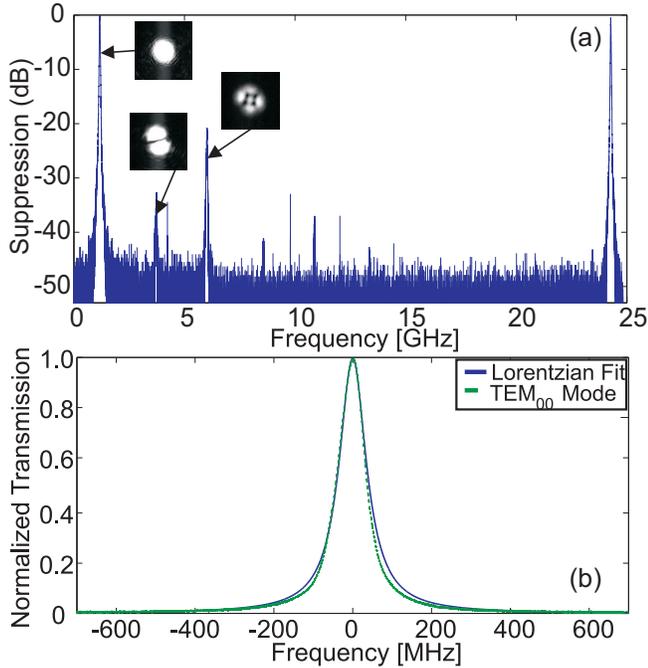}}
\caption{(a) The transmission of the cavity over a FSR. The transverse intensity profile of the first three eigenmodes is shown in the inset. (b) The transmission profile of the TEM$_{00}$ mode with green line showing the experimental data and the blue showing the theoretical fit. The FWHM is found to be 84 MHz}
\label{FSR}
\end{figure}

\noindent \textbf{Construction and Design}: Our filter cavity consists of a commercial plano-convex lens from Lambda Research Optics with a surface quality of $\lambda/10$ at 633 nm, a scratch/dig of $10/5$, and high reflectivity coating ($R=99.0\%\pm0.1\%$) on each surface. For the purpose of our experiments on Raman-like effects in atomic rubidium \cite{Andrew}, where we filter single photons from a strong pump, separated in frequency by $3$ to $7$ GHz, we choose substrates with centre thicknesses $L=4.3$, $5.3$, and $7.8$ mm. The curvature radius is chosen as $r=40.7$ mm, which satisfies the stability condition for plano-convex lenses: $0<1-L/r<1$. We choose BK7 as the substrate material owing to its high transmission in the near infrared and its high coefficient of thermal expansion.

The cavity is placed in a standard lens mount with a thermally coupled AD$590$ sensor measuring the temperature to within 0.1 $^{\circ} \text{C}$. Temperature control is achieved through a Peltier thermoelectric cooler which couples the mount to a large aluminum block which acts as a heat sink. The temperature sensor and Peltier element are connected to a standard PID temperature control system (Thorlabs ICT100) with a long-term stability of 0.004 $^{\circ} \text{C}$. The entire system is enclosed to minimize environmental coupling. The optical field for probing the cavity is provided by a continuous-wave Ti:Sapphire laser aligned to match the TEM$_{00}$ mode of the cavity.

\noindent\textbf{Results}: We characterize the spectral properties of the cavity by scanning the laser over one cavity FSR and monitoring the transmission as shown in Fig.~1. For the $4.3$ mm cavity, we measure $\text{FSR}=23.1\pm0.2$ GHz which is consistent with the expected value of $c/2nL=23.07$ GHz. Since the laser linewidth ($100$ kHz) is much less than the cavity linewidth we neglect its contribution and find $\Delta\nu=84 \pm 5$ MHz [Fig.~1(b)] corresponding to a finesse of $\mathcal{F}=275 \pm 19$, which is somewhat less than the value of $\mathcal{F}=312\pm 30$, expected according to \eqref{finesseEq}, due to various defects as discussed above. 

The on-resonance transmission reaches $\mathcal{T}_{\max} \approx 0.6$, indicating that the mirror defects and losses do not play a significant role in the cavity performance. For the spectral component located $6.8$ GHz away from the cavity resonance, we find an extinction ratio of over $45$dB, in agreement with Eq.~\eqref{exratEq}. Given that the intracavity losses are low in this configuration, at least a 10 dB higher extinction ratio can be obtained with a similar lens by choosing a higher reflectivity coating. Further improvements can be achieved by using a substrate lens with superior surface characteristics.

Even with the present cavity, the extinction ratio can be greatly improved by placing a spatial filter at the output and if polarization filtering is available. Acting in this fashion, in a separate experiment, we achieved over $180$ dB suppression of a strong pump beam, tuned $\approx3$ GHz from the resonance \cite{Andrew}. Thus our design provides an attractive compromise between simultaneous high transmission of the desired signal and attenuation of unwanted modes.

A key figure of merit in our filter is the suppression of unwanted spectral and spatial modes. Imperfect mode-matching results in the appearance of higher order transverse modes which cause intermittent resonance peaks along the spectrum [Fig.~1(a)] and thus lowers the suppression. However, these peaks are sparse and for a broad range of frequencies do not significantly degrade the extinction. The distance between adjacent peaks is 2.4 GHz, which is consistent with Eq.~\eqref{transsepEq}.

\begin{figure}[htb]
\centerline{\includegraphics[width=8.5cm]{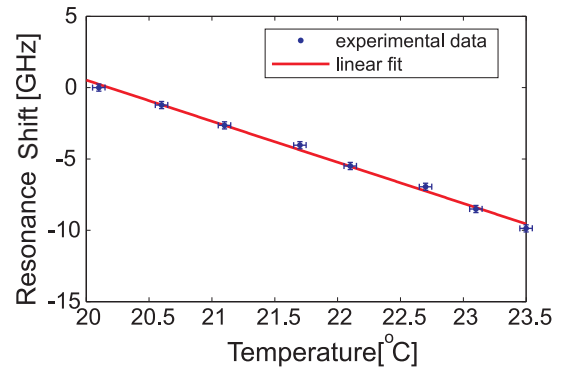}}
\caption{Temperature tuning of the resonant frequency. The frequency change of TEM$_{00}$ mode is displayed, which is linear within 4 K with the slope of $-$2.88 GHz/K.}
\label{TEM00}
\end{figure}

In order to characterize the temperature tunability of the system, we monitor the deviation of the $\text{TEM}_{00}$ transmission maximum as a function of temperature as shown in Fig.~\ref{TEM00}. The behaviour is seen to be linear with slope $d\nu/dT=-2.88$ GHz/K. The calculation gives $d\nu/dT=-3.32$ GHz. The discrepancy is most likely due to inaccuracy in the stated values for BK7, which vary by up to $20\%$ depending on the source. 

\begin{figure}[htb]
\centerline{\includegraphics[width=8.5cm]{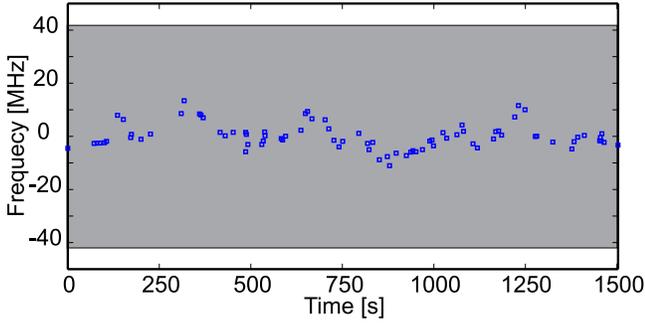}}
\caption{Long term drift of cavity resonance frequency. The shaded area represents the cavity line-width. We observe a drift of $0.057\Delta\nu$ over the 25 minute interval.}
\label{drift}
\end{figure}

The long term stability of the cavity system depends largely on the accuracy of the temperature-control system. To characterize this parameter, we observe the frequency deviation of the transmission maximum with respect to a local oscillator which is stabilized to an atomic transition. We observe a rms drift of $0.057\Delta\nu$ for data taken over 25 minutes and $0.0950\Delta\nu$ for 2 hours. The latter value corresponds to an rms temperature drift of 0.003 K, indicating that the temperature controller is likely the limiting factor in the cavity stability. Figure \ref{drift} shows the measured frequency fluctuations over a 25-minute span.

\begin{figure}[htb]
\centerline{\includegraphics[width=8.5cm]{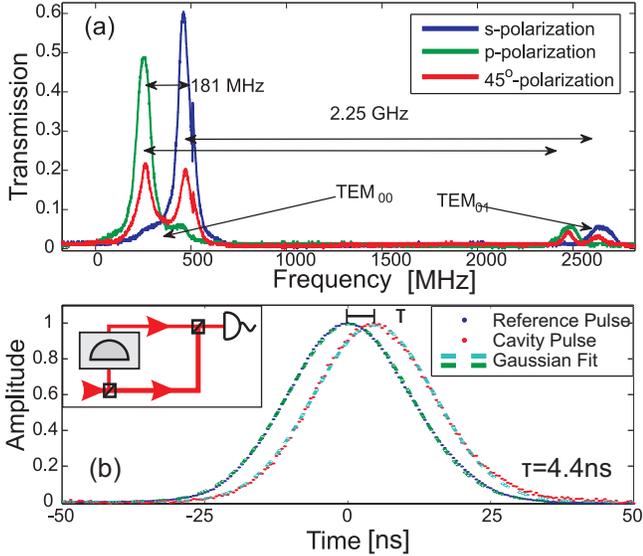}}
\caption{(a) The transmission profile of different polarizations through the 5.3-mm cavity showing a frequency separation of 181 MHz between the resonances for the $s$ and $p$ polarizations. (b) The delay of a laser pulse propagating through the cavity is found to be 4.40 ns. The signal is obtained from a setup shown in the inset schematic.}
\label{birefringence}
\end{figure}

Since even minor differences in the path length cause significant shifts in the resonant frequency, any birefringence in the cavity material will cause separate peaks for the ordinary and extraordinary polarizations. As shown in Fig. 4(a), the $s$ and $p$ polarization resonances are separated by $\sim 181$ MHz for this particular cavity.

Another figure of interest for time-domain experiments is the delay introduced by our filter cavity due to its on-resonance dispersion. To determine this, we measure the delay of a 40 ns pulse with respect to a reference pulse  [Fig.~\ref{birefringence}(b)] and observe a value of 4.4 ns. A calculation using the Kramers-Kronig relations yields the effective group velocity to be $1.04\times 10^6$ m/s which corresponds to a 5.3 ns delay. The discrepancy with the experimental result can be attributed to uncertainties in the measurements of the arrival time or approximations used in the group velocity calculation.

\noindent\textbf{Summary}: We have built and characterized a monolithic filter cavity for use in quantum optics experiments. The cavity consists of a simple plano-convex lens subjected to dielectric high-reflection coating on both sides, and is tuned by changing its temperature. In a non-confocal geometry, it is possible to obtain stability and transverse mode filtering as well as high extinction ratio in a single-pass configuration.

The work was supported by NSERC and CIFAR. We thank Irina Novikova and Ben Buchler for helpful discussions.

\bibliographystyle{unsrt}

\begin{thebibliography}{99}
\bibitem{Polzik} J. S. Neergaard-Nielsen \textit{et al.,} Phys. Rev. Lett. \textbf{97,} 083604 (2006).
\bibitem{Furusawa} N. Lee, \textit{et al.,} Science \textbf{332,} 330 (2011).
\bibitem{Andrew} A. MacRae, T. Brannan, and A. I. Lvovsky, {\tt arXiv:1112.4855 [quant-ph]} (2012).
\bibitem{Gem_11} M.~Hosseini, \textit{et al.} Nature Physics \textbf{7} 794--798 (2011).
\bibitem{vanderwal03} C. H. van der Wal, \textit{et.al.}, Science \textbf{301}, 196 (2003).
\bibitem{Roy-Hercher} C. Roychoudhuri and M. Hercher, Appl. Opt. \textbf{16,} 2514 (1977).
\bibitem{Sloggett_84} G.~J.~Sloggett Appl. Optics, \textbf{24}, 14 2427--2432 (1984)
\bibitem{Benson} D. H\"ockel, E. Martin, and Oliver Benson, Rev. Sci. Instrum. \textbf{81}, 026108 (2010).
\bibitem{Hernandez} G. Hernandez, ``Luminosity and resolution considerations,'' in \textit{Fabry-Perot Interferometers} (Cambridge University Press, 1986), pp. 119-172.
\bibitem{maunz2004} P. Maunz, \textit{et.al.}, Nature \textbf{428}, 50-52 (2004).
\bibitem{hood2001} C. J. Hood, H. J. Kimble, Jun Ye, Phys. Rev. A \textbf{64}, 033804 (2001).
\bibitem{Hecht} Eugene Hecht, ``Interference'' in \textit{Optics} 4th ed. (Addison-Wesley, 2002), pp.385-438.
\bibitem{Hercher_68} M. Hercher, Appl. Opt. \textbf{7}, 5 951--966 (1968).
\bibitem{Li} H. Kogelnik and T. Li, Proc. IEEE \textbf{54,} 1312 (1966).
\bibitem{Ghosh} G. Ghosh, Appl. Opt. \textbf{36} 7 1540--1546.


\end{thebibliography}

\end{document}